**Vacuum Rabi splitting in a plasmonic cavity at the single quantum emitter limit**

Kotni Santhosh[1,*,#], Ora Bitton[2,*], Lev Chuntonov[3,*] and Gilad Haran[1]

*[1]Departments of Chemical Physics and [2]Chemical Research Support, Weizmann Institute of Science, Rehovot, Israel, and [3]Schulich Faculty of Chemistry, Technion-Israel Institute of Technology, Haifa, Israel.*

Correspondence and requests for materials should be addressed to G.H. (gilad.haran@weizmann.ac.il).

*These authors contributed equally to this work.

# Current address: Hindustan Petroleum Green R&D Center, HPCL, Bangalore, India.

**Abstract**

**The strong interaction of individual quantum emitters with resonant cavities is of fundamental interest for understanding light matter interactions. Plasmonic cavities hold the promise of attaining the strong coupling regime even under ambient conditions and within subdiffraction volumes. Recent experiments revealed strong coupling between individual plasmonic structures and multiple organic molecules, but so far strong coupling at the limit of a single quantum emitter has not been reported. Here we demonstrate vacuum Rabi splitting, a manifestation of strong coupling, using silver bowtie plasmonic cavities loaded with semiconductor quantum dots (QDs). A transparency dip is observed in the scattering spectra of individual bowties with one to a few QDs, directly observed in their gaps. A coupling rate as high as 120 meV is registered even with a single QD, placing the bowtie-QD constructs close to the strong coupling regime. These observations are verified by polarization-dependent experiments and validated by electromagnetic calculations.**

**Introduction**

The interaction of emitters with an optical cavity falls within the realm of cavity quantum electrodynamics[1]. Approaching the limit of strong coupling between individual quantum emitters and resonant cavities is important for multiple advanced optical applications, such as quantum information processing[2,3] and quantum communications[4,5]. Strong coupling can be observed through the phenomenon of vacuum Rabi splitting in the optical spectra of the joint systems, as well as in the appearance of nonclassical photon correlations. The strength of the interaction depends on the ratio of the quality factor of the cavity to the mode volume, $Q/V$. In optical cavities made of e.g. photonic crystals[6] or micropillars[7] the fundamental laws of diffraction set a strict limitation on the cavity size, which cannot be smaller than half the wavelength of the interacting photon. This in turn limits how small $V$ can be, and mandates a very high $Q$ in order to attain the strong coupling regime. Obtaining such a high quality factor requires demanding experimental conditions such as cryogenic temperatures and ultra-narrow frequency light sources.

Cavities made of materials that can sustain surface plasmon (SP) excitations can beat the diffraction limit by focusing intense electromagnetic fields into volumes much smaller than the wavelength of light[8]. Such cavities should simplify considerably the experimental conditions required for strong light-matter interactions, and may allow quantum optical experiments to be conducted under ambient conditions.

Noble metals are most often used for generating structures sustaining SPs at visible wavelengths. While the SP relaxation times in such structures are ultrafast, severely limiting $Q$, their mode volume is drastically reduced compared to photonic cavities. In recent years there has been much interest in studies of coupling between surface plasmons, either propagating or localized, and quantum emitters such as molecules[9]. These experiments involved multiple molecules and often also multiple or extended plasmonic structures.

Recently, strong coupling has also been probed at the level of a single plasmonic device. Thus, Halas and coworkers used gold dimers and molecular J-aggregates to observe a Rabi splitting of ~230 meV[10]. Shegai and coworkers observed strong coupling using either a silver nanorod[11] or a silver nanoprism[12], in both cases employing J-aggregates as the quantum emitters. In all of these experiments the interaction involved hundreds of quantum emitters or more. However, for quantum information operations one needs to approach *the limit of a single quantum emitter coupled to the cavity*.

In this work we show that one can indeed observe strong coupling in the limit of a single quantum emitter. In particular, we use silver bowtie plasmonic cavities and couple them to semiconductor quantum dots (QDs). Scattering spectra registered from individual plasmonic cavities containing one to a few QDs show vacuum Rabi splitting, indicating that the strong coupling regime is approached in these systems. Polarization-dependent experiments verify that the observed Rabi splitting is due to the coupling of the longitudinal plasmon resonance of the bowties with the QDs.

**Results**

**Construction of plasmonic cavities with QDs.** Inspired by recent computational studies suggesting that strong coupling with individual emitters can be achieved in the gaps of plasmonic structures[13], we turned to silver bowties, which can be fabricated with a small gap[14]. Silver is preferred as a material for plasmonic devices due to its relatively low surface plasmon damping and hence the high achievable quality factors of such devices compared to other metals[15]. By varying the structural parameters of the bowties, including side length and gap size, the localized surface plasmon excitations of bowties manufactured by electron-beam lithography can be tuned to resonance with the quantum emitter of choice, while maintaining gaps of the order of ~20 nm. To achieve strong coupling the quantum emitter should possess a large oscillator strength and a narrow linewidth. QDs have several advantages over organic emitters in relation to the aforementioned characteristics[16], and (most importantly) are also significantly more photostable. They can also be observed by electron microscopy and therefore can be directly counted. Commercially-available CdSe/ZnS QDs were employed in the present study. Optical spectroscopy and electron microscopy studies showed that the sample is mildly heterogeneous and the diameter of the QD varies between 6 and 8 nm.

To position QDs within the gaps of bowties we made use of interfacial capillary forces[17] to drive the QDs into lithographically patterned holes in the bowtie gaps. This simple method avoids additional steps such as chemical modification of the QD/nanostructure surfaces. The process, described in Figure 1 and in the Methods section, led to the trapping in bowtie gaps of a number of QDs that varied from one to several, as could be ascertained by scanning electron microscopy.

**Spectroscopy demonstrates Rabi splitting.** Dark field microspectroscopy was used in order to characterize the plasmonic behavior of every single bowtie by measuring its scattering spectrum. As shown in Supplementary Figure 1, empty bowties possess two modes, a transverse mode at ≈ 2.05 eV and a longitudinal mode at ≈ 1.9 eV, the latter being due to dipolar coupling of the two parts of each bowtie. The absorption and emission spectra of the QDs are shown in Supplementary Figure 2, and demonstrate that the optical transitions of the QDs are in resonance with the plasmon excitations of the bowties. (See also Supplementary Note for further discussion on the role of the absorption and emission spectra in coupling to the plasmonic cavity.)

Figure 2a shows bowties with QDs in their gaps whose scattering spectra present transparency dips indicative of Rabi splitting. Overall we collected scattering spectra from 21

QD bowties, out of which 14 spectra showed a definite signature of splitting. Thus, the top bowtie in Figure 2a, with a single QD and a particularly small gap (~19 nm), presents a clear Rabi splitting feature. Similarly, Rabi splitting is observed in the spectra of the two other bowties shown in Figure 2a, with two and three QDs and gaps of 17 and 30 nm, respectively. The spectra in this figure, as well as other spectra showing Rabi splitting, were fitted using the coupled oscillator model[18] (see Methods). This model represents the plasmon mode and the exciton of the quantum emitter as coupled harmonic oscillators exchanging energy reversibly. The coupling rate characterizing the interaction is given by $g = \mu \cdot E$, where $\mu$ is the transition dipole moment of the exciton and E is the electric field of the plasmon. This coupling leads to the generation of upper ($\Omega_+$) and lower ($\Omega_-$) plasmon-exciton hybrid states, with a transparency dip in between[13,19]. The Rabi splitting values, $\Omega_+ - \Omega_-$, can be obtained directly from the fitted curves, and for the three examples in Figure 2a they are 176, 288 and 224 meV, respectively. These values are comparable to recently reported values of Rabi splittings of J-aggregates with individual plasmonic structures[10,12], although hundreds of quantum emitters were involved in the latter. Not all bowties loaded with a single QD showed a clear transparency dip in the scattering spectrum. Instead, Rabi splitting manifested itself in a significant spectral broadening as compared to that of an empty bowtie (Supplementary Figure 3), indicating coupling between plasmons and excitons. Splitting was readily observed in bowties with several QDs- see Supplementary Figure 4 for additional spectra.

Coupling rates obtained from the fits of data sets with one and two QDs are plotted as a function of gap size in Figures 2b-c. The figures also contain theoretical estimates of the coupling rates, based on the numerical calculations described below. The coupling rate depends strongly on the gap size of each bowtie. For instance, for two QDs the coupling rate is more than doubled when the gap size changes from 25 to 17 nm. In general, the coupling increases with the number of quantum emitters (N) ($g \approx \sqrt{N} g_1$, $g_1$ being the coupling rate for one quantum emitter[1]), though in our experiments this dependence is partially masked by the dependence on gap size and QD position variations. Though the values of the coupling rates obtained from the fits are smaller than the values of the plasmon widths (which are ~400 meV), significant splittings are observed, indicating that the strong coupling regime has indeed been attained, or is close[9]. Supplementary Figure 5 shows histograms of the values of all parameters obtained from the coupled oscillator fits to measured spectra.

To obtain an unequivocal proof that the splitting in the spectra is due to strong (or close-to-strong) coupling of the bowtie longitudinal plasmon mode and the quantum emitter's exciton, we resorted to polarization-dependent experiments, in which the polarization of the excitation source was rotated. Figure 2d shows an example of a polarization series measured on the bowtie with two QDs from Figure 2a. When the excitation is polarized along the longitudinal direction of the bowtie (i.e. along its long axis) the spectrum shows two peaks due to Rabi splitting. As the polarization is rotated to the transverse direction the double-peaked spectrum is gradually replaced with the single-peaked spectrum of the transverse mode. Similar results with other bowties are shown in Supplementary Figure 6. These results clearly indicate that the transparency dips in the scattering spectra are due to a genuine coupling of the plasmon and QD excitations.

**Discussion**

The experimental results described above show that the positioning of one or a few QDs within a silver bowtie can bring this plasmonic cavity very close to the strong coupling regime. In order to support the experimental results and shed further light on them, we performed electromagnetic calculations, using the boundary element method[20]. The simulations revealed some interesting facts. The top spectrum in Figure 3a is the calculated scattering spectrum of an empty bowtie. The spectrum for a bowtie with a single QD at the center of the gap is shown below, demonstrating a small change compared to the empty bowtie. However, when the QD is positioned closer to one of the prisms constituting the bowtie, or the number of QDs in the hotspot is increased from one to two, a more significant transparency dip appears. The Rabi splitting is observed in both calculated scattering and extinction spectra (the latter not shown). Importantly, no such splitting is found when the simulation is repeated with a single metallic prism rather than a bowtie structure.

The simulation allows us to directly calculate the coupling rate for a quantum emitter with an oscillator strength similar to a QD (~0.6)[21-23]. The spatial distribution of the coupling rate across the gap is shown in Figure 3b, clearly indicating that the electromagnetic field in the hot spot is not uniform and is concentrated at the edges rather than at the center. Indeed, in many of the bowties exhibiting strong coupling (Figure 2 and Supplementary Figures 3 and 4) the QDs are close to one of the edges, in agreement with the simulation results. We estimate from the simulation that at the center of a bowtie structure $g$ is ~100 meV, while very close to one of the prisms $g$ is as high as 200 meV.

In conclusion, we have successfully integrated lithographically fabricated silver bowties with one to a few QDs that reside exactly within the plasmonic cavity. The very small plasmon mode volumes of the bowties allowed us to demonstrate vacuum Rabi splitting in the limit of an individual quantum emitter. A key advantage of using QDs is that we can directly count them under the electron microscope and verify the relation between coupling and the number of quantum emitters within each cavity. An important question to ask is: what value of the coupling may deem it 'strong'? An often-used rule of thumb for the strong coupling regime is $2g > (\gamma_0 + \gamma_{pl})/2$, where $\gamma_0$ and $\gamma_{pl}$ are the QD and plasmonic cavity linewidths, respectively [9,24]. The maximal coupling rates obtained in this work for two and three QDs (~200 meV, Figure 2b-c and Supplementary Figure 5d) fulfil the above criterion, while the maximum value obtained for a single QD (120 meV) is somewhat smaller. This shows that the constructs of QDs with plasmonic cavities studied here are definitely close to the strong coupling regime. In future experiments we plan to significantly increase the coupling between individual QDs and bowties. To this end we will reduce the mode volume of the bowties, which can be achieved by decreasing their gap size and improving control over their shape in the fabrication process. Further improvement in QD positioning methodology will allow us to locate a single QD reproducibly at the position of the highest field within a bowtie cavity. Such improvements will enable systematic variation of experimental parameters and therefore provide additional proofs for strong coupling. Importantly, our results also pave the way to more sophisticated quantum nonlinear optics experiments[25]. In particular, by probing time-resolved photoluminescence of the coupled QD-bowtie system we hope to directly demonstrate quantum correlations [26,27].

**Methods**

**Materials.** Glass/ITO (10 mm × 10 mm × 1.1 mm) substrates were purchased from Xin Yan Technology Ltd., China. The ITO layer is 180 nm thick, and is characterized by 83 % transparency to visible light and a sheet resistance of 10 ohm/sq. Poly(methyl methacrylate) 950K A2 (PMMA) electron beam resist, used for the fabrication of bowties, was procured from MicroChem. Chromium (used as an adhesive layer) and silver were obtained from Kurt J. Lesker. Water soluble, mercapto-undecanoic acid capped CdSe/ZnS core/shell nanocrystals (quantum dots) were acquired from MK Impex Corp.

**Electron Beam Lithography and Fabrication of Silver Bowties.** Prior to fabrication, Glass/ITO substrates were cleaned with acetone followed by isopropanol for 3 minutes in each step. The cleaned glasses were dried using a nitrogen flow and then PMMA was spin-coated at a speed of 6000 rpm for 50 sec to achieve a thickness of 60 nm, followed by baking at 180°C for 90 sec. The PMMA-coated glasses were loaded into a Raith E_line Plus electron beam lithography system and the PMMA was exposed to define the shape of bowties. The accelerating voltage used for the exposure was 30 kV and the beam current was 30 pA. The design consisted of matrices of bowties, with each matrix hosting 64 bowties. Each bowtie was separated by 10 μm from its neighboring partner to guarantee no interaction. To remove the exposed PMMA, the substrates were developed using 1:3 methyl isobutyl ketone:isopropyl alcohol for 30 sec, followed by immersion in a stopper (isopropyl alcohol) for 30 seconds and drying with a nitrogen flow.

An electron beam evaporator (Odem) was used for the metal deposition on the patterned substrates. Initially chromium was evaporated to deposit a 2 nm adhesion layer, followed by silver. Different prism lengths from 75 to 200 nm and thicknesses from 20 to 35 nm were examined and finally the dimensions were fixed to a length of ≈ 85 nm and a thickness of ≈ 30 nm. These bowties showed a plasmon resonance that overlaped the QD exciton frequency. After metallization a liftoff step was performed in acetone for 5 minutes to obtain silver bowties on the glass/ITO substrate.

A second stage of E-beam lithography was performed in order to expose holes in the gap regions of the bowties. A 60 nm layer of PMMA was spin coated on the substrate and then the gap regions of the bowties were exposed to create 10-30 nm holes, into which the QDs were driven. (See below and in Figure 1 of the main text.) In order to reach an overlay accuracy as high as a few nanometers in the second exposure, we made use of alignment marks that were fabricated in the first E-beam exposure step.

**Incorporation of QDs into bowtie structures.** To incorporate QDs into the bowtie gaps we adopted a method originally developed by Alivisatos and coworkers[17]. Substrates with bowties and holes in the PMMA layer prepared as described above were placed vertically in an aqueous solution of QDs. Controlled solvent evaporation was then used to exert a capillary force along the receding line of contact of the QD solution and drive QDs successfully into the holes. By varying the QD concentration in solution (from 10 to 100 nM), bowties with one to a few QDs in the gap were successfully obtained (Figure 1). Though this method is limited in controlling the exact position of the QDs within the gap, it was found to be facile and relatively reproducible.

**Bulk Absorption and Emission Measurements of QDs.** Steady state absorption and emission spectral measurements (Supplementary Figure 2) were carried out on a Cary-100 UV-VIS spectrophotometer (Varian) and a Fluorolog-3 spectrofluorimeter (Horiba Jobin Yvon), respectively. Extinction coefficients of the QDs were estimated by employing the method developed by Peng and coworkers[23].

**Dark-field Microspectroscopy.** Our dark-field microspectrometer is described schematically in Supplementary Figure 8. Briefly, scattering spectra of single bowties, either with or without QDs, were measured with an inverted microscope equipped with a dark field condenser, a 75 W Xenon lamp (Olympus) and a 100× oil immersion objective with NA = 1.30. A SpectraPro-150 spectrograph with a 1200 g/mm grating (Acton) and a Newton spectroscopy CCD camera (Andor Technology) were used to disperse scattered photons and register spectra. Raw spectra were smoothed using a Fourier low-pass filter. Prior to the spectral measurements the camera was calibrated in reference to the spectrum of fluorescent dyes. Polarization of the excitation light was controlled by a combination of a polarizer and a selecting sector, such that essentially only s-polarized light reached the sample.

**Data fitting to the coupled-oscillator model.** Scattering spectra were fitted to the following equation, which represents the scattering of two coupled harmonic oscillators at frequency $\omega$ [18]:

$$\sigma_{\text{sca}}(\omega) = A\omega^4 \left| \frac{(\omega^2 - \omega_0^2 + i\omega\gamma_0)}{(\omega^2 - \omega_0^2 + i\omega\gamma_0)(\omega^2 - \omega_{\text{pl}}^2 + i\omega\gamma_{\text{pl}}) - \omega^2 g^2} \right|^2 \quad \ldots\ldots\ldots\ldots (1)$$

In this expression $\omega_{\text{pl}}$ and $\omega_0$ are the plasmon and QD resonance frequencies, respectively, while $\gamma_{\text{pl}}$ and $\gamma_0$ are the corresponding linewidths. $g$ is the coupling rate and $A$ is a scaling parameter. Parameters obtained from fits to all our data sets are shown in Supplementary Figure 5. We fixed the value of $\gamma_0$ to 130 meV, which is the measured linewidth of the fluorescence spectra of individual QDs (not shown). In the case of some of the data sets with a single QD in the gap, where broadening of the spectrum was observed, though no clear splitting, we also fixed the value of $\gamma_{\text{pl}}$ to 385 meV, which is the average value of this parameter obtained from multiple spectra of empty bowties.

**Electromagnetic Simulations.** Numerical simulations of the silver bowties were performed using the boundary-element method as implemented in MNPBEM – a toolbox developed by Hochenester and co-workers.[20] A numerical model of a bowtie was constructed from two prisms with equilateral triangular bases of 80 nm side length, 30 nm heights, radii of curvature at the vertices of 7 nm, and a gap size of 17 nm. The complex refractive index of silver from Johnson and Christy [28] was used, and a refractive index of 1.33 was assumed for the ambient medium. The bowtie was excited by a plane wave at normal incidence and polarized along the longitudinal direction and the plasmon scattering spectra were calculated at the far field. From the spectrum of an unloaded cavity we calculated $\gamma_{\text{pl}}$=0.25 eV, which is smaller than the typical experimental linewidth. However, the cavity quality factor: $Q = \omega_{\text{pl}}/\gamma_{\text{pl}}$=7.3, was only slightly higher than the measured quality factor. Simulations of scattering spectra for a bowtie loaded with QDs were performed with dots modeled as spheres of 8 nm located at various positions within the gap. The complex dielectric function of these dots was approximated by a Lorentz model with a high-frequency dielectric constant $\epsilon_\infty$=6.1, $\omega_0$=1.8 eV, $\gamma_0$=0.08 eV, and an oscillator strength $f$=0.6[21-23]. Since the computed plasmon linewidth of the idealized bowtie structure was consistently smaller than the measured linewidth of a real bowtie, we scaled the linewidth of the QD accordingly.

The position-dependent coupling rate of a QD strongly coupled to the bowtie $g(r)$ was evaluated using $E(r)$, the electric field numerically calculated at the plasmon resonance energy $\omega_{\text{pl}}$=1.83 eV. To this end, the mode volume of the antenna was calculated using the approach of Koenderink [29], and was used to normalize the energy density within the cavity, assuming it is occupied with a single photon of energy 1.83 eV. A map of $g(r)$ plotted on a horizontal plane that intersects the bowtie at its half-height is shown in Figure 3b. Supplementary Fig. 7 shows a similar calculation for a single prism (panel a), and compares the coupling rate of a bowtie and a single prism (panel b).

**Data availability.** All relevant data are available from the authors upon request

**Acknowledgements**

GH is the incumbent of the Hilda Pomeraniec professorial chair. We thank Dr. Barak Dayan for useful comments on the manuscript.

**Author Contributions**

SK, OB and GH designed the research. SK and OB performed the experiments. LC performed electromagnetic simulations. All authors participated in writing the manuscript.

**Competing financial interests**

The authors declare no competing financial interests.

**Figure legends**

**Figure 1. Construction of bowties with quantum dots in their gaps. a**, Schematic illustration of the two-step lithography process for making holes at the center of bowtie structures and the interfacial capillary force assisted method for driving QDs into the holes. **b**, Scanning electron microscope images of bowties with one, two and multiple QDs in their gaps (from top to bottom). The positions of the QDs are marked by red arrows. Yellow scale bars represent 20 nm.

**Figure 2. Strong coupling of plasmons and quantum emitters. a**, Scattering spectra of bowties with (from top to bottom) one, two and three QDs in the gap, respectively. All spectra show a transparency dip due to Rabi splitting. The black lines are experimental data, and the colored lines are fits to the coupled oscillator model. Insets show the scanning electron microscope images of the bowties. The positions of the QDs are marked by red arrows. Yellow scale bars represent 20 nm. **b-c**, Coupling rates as a function of gap size for bowties with one QD (**b**, red symbols) and two QDs (**c**, green symbols). The errors in the coupling rate values, obtained from the fitted functions, are estimated to be 2-5 meV. The continuous lines represent the numerically calculated coupling rates at two configurations along the center line of the bowtie: with the QDs almost touching one of the prisms (continuous lines) or with the QDs at the center of the bowtie (dashed-dotted lines). In the experiments the QDs may be positioned away from the center line, so that their coupling rates are lower than the calculated lines. **d**, Polarization series of the middle bowtie structure in **a**. As the polarization of the excitation light is rotated from the direction parallel to the bowtie long axis to perpendicular to it, the transparency dip in the spectrum gradually vanishes, indicating that it indeed originates in the coupling of the QD exciton with the longitudinal plasmon resonance of the bowtie.

**Figure 3. Electromagnetic simulations of strongly-coupled plasmonic structures. a**, Simulated scattering spectra of bowties with one and two QDs organized as indicated in the inset structures. **b**, Distribution of the coupling rate (in eV) of a quantum emitter with an oscillator strength of 0.6 in a bowtie structure. The white bar represents 10 nm.

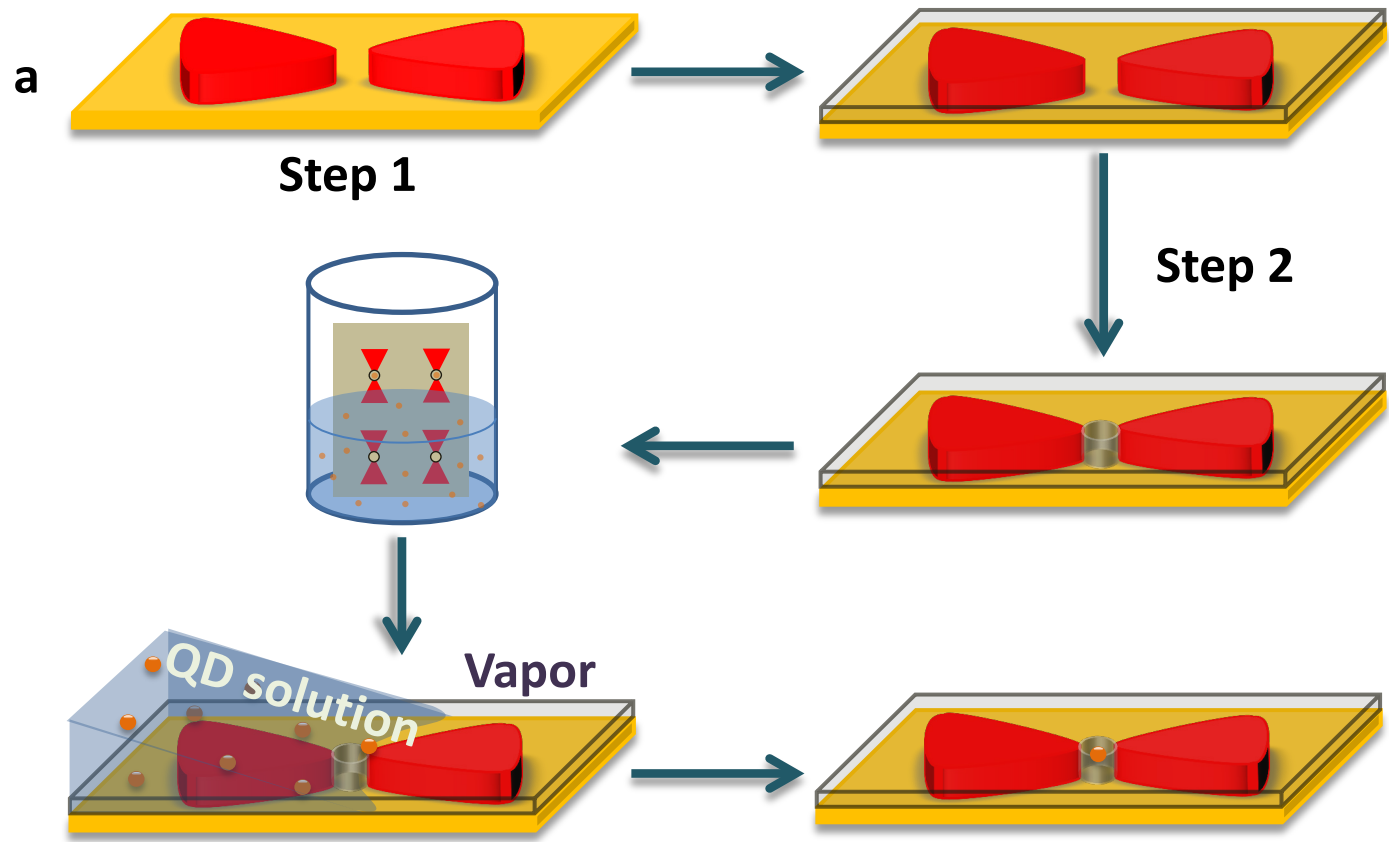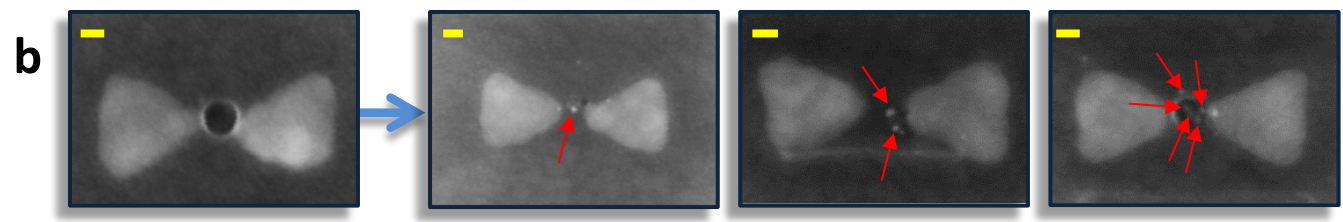

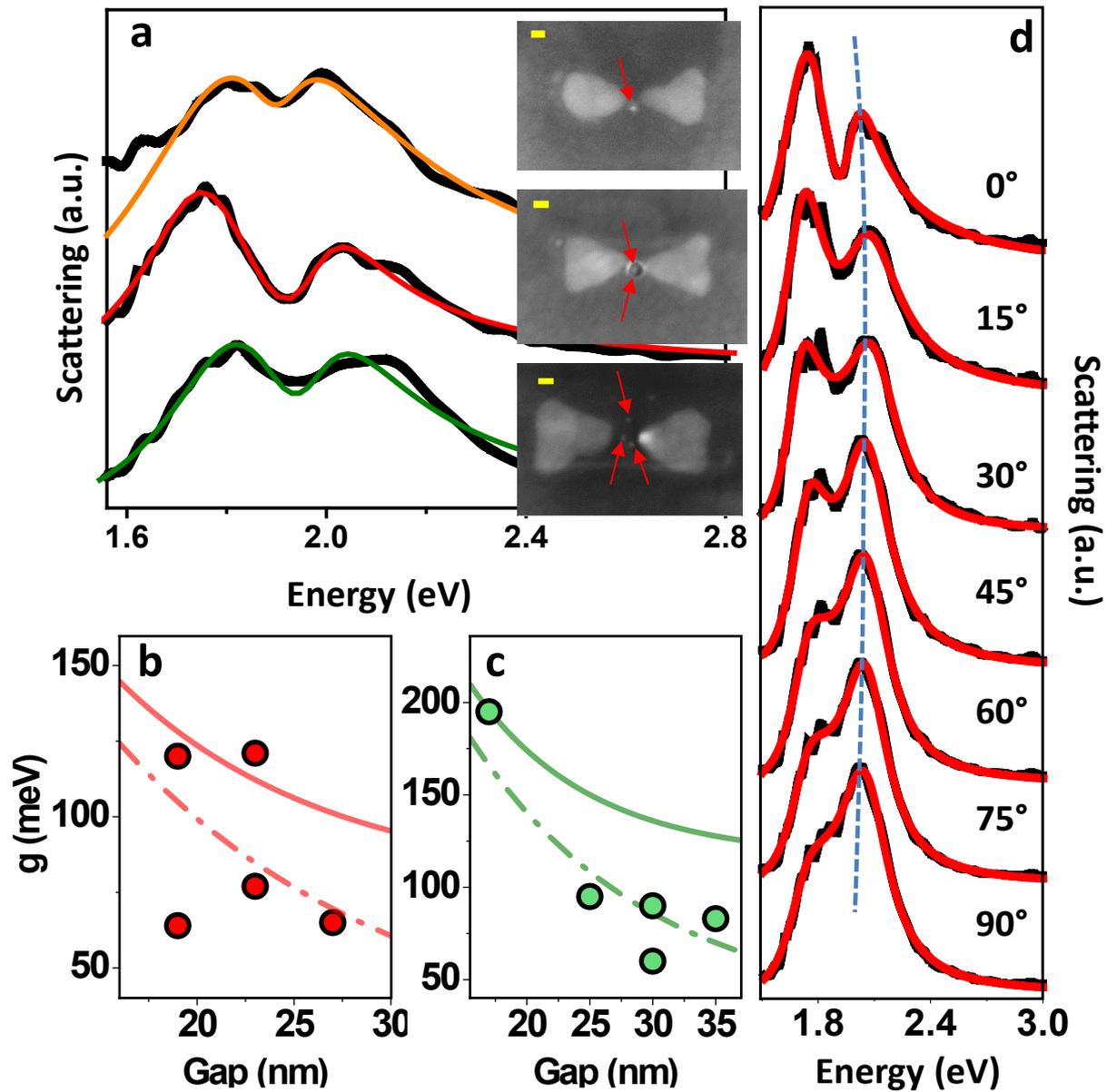

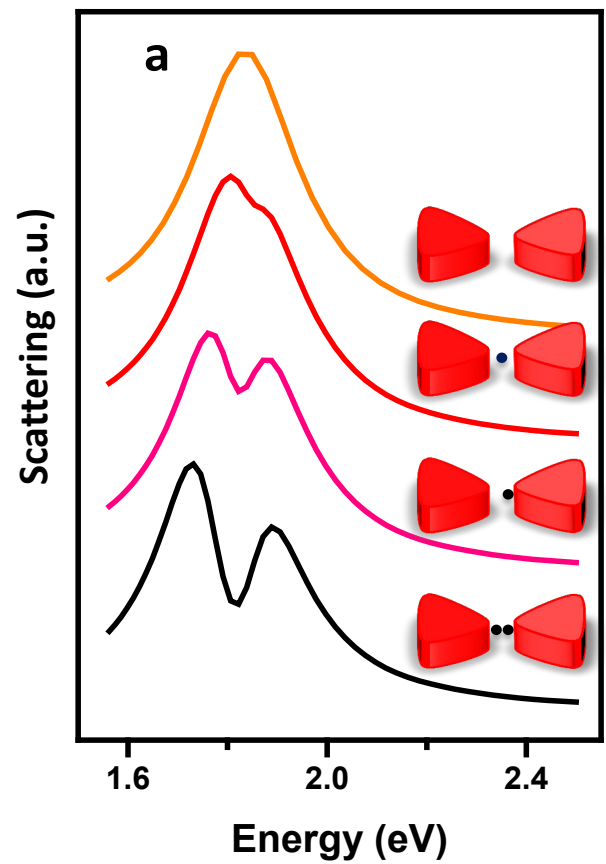
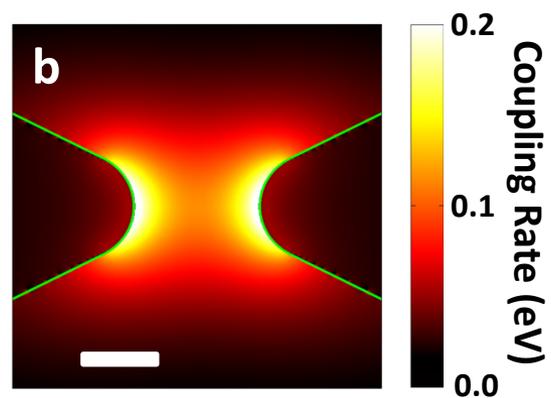

# Supplementary Information

**Supplementary Figure 1**

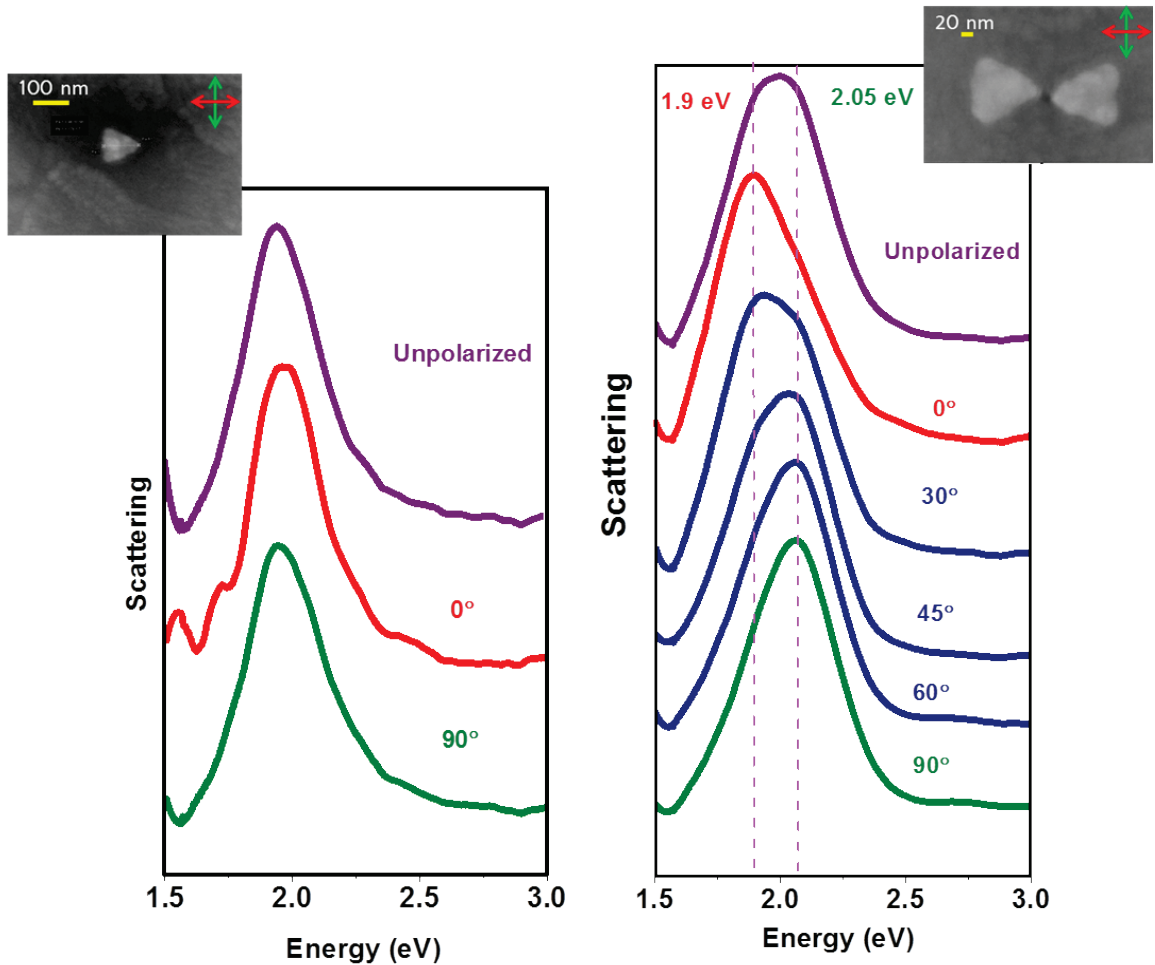

**Dark field scattering spectra of an individual prism and an individual bowtie as a function of angle of polarization.** The corresponding electron microscope images of the structures are also shown. Red and green arrows denote the direction of polarization parallel to the bowtie axis and perpendicular to it, respectively. The solid lines are smoothed versions of the data.

**Supplementary Figure 2**

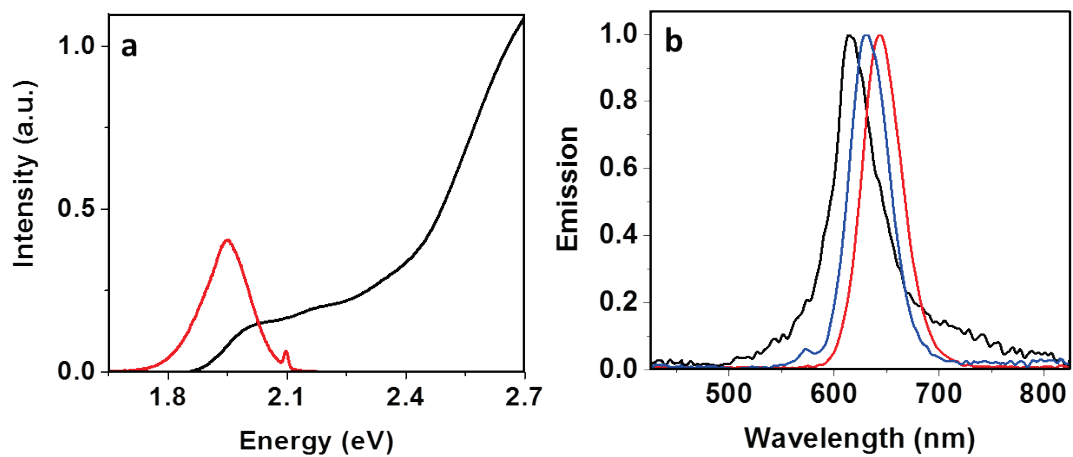

**Spectroscopy of quantum dots in water. a.** The figure shows absorption (black) and emission (red) spectra of CdSe/ZnS QDs in water. The lowest energy optical transition of the QDs (at 1.9 eV ≈ 660 nm) is at resonance with the longitudinal plasmon excitation seen in Supplementary Figure 1. **b.** Photoluminescence spectra of individual QDs deposited on a glass surface and excited with a 532 nm laser. Variations in the position of the luminescence peak are ascribed to heterogeneities in QD sizes. For a comment on the effect of the shape of the spectra of QDs on strong coupling see Supplementary Note.

**Supplementary Figure 3**

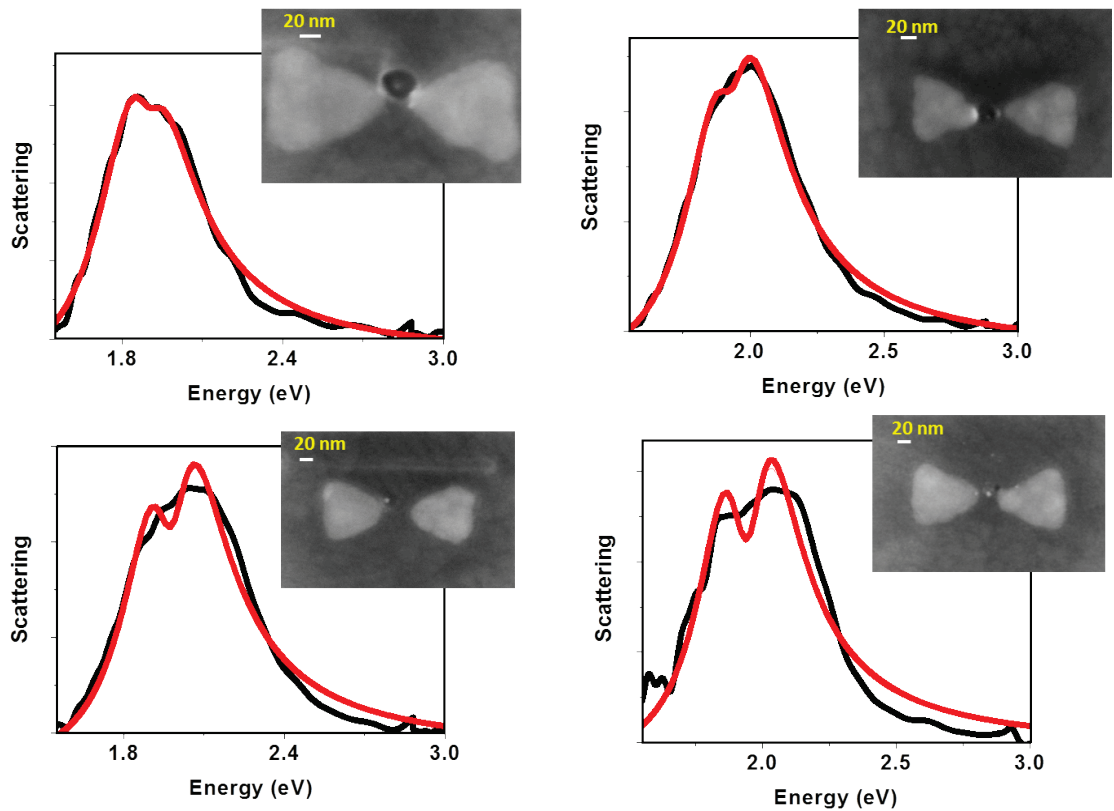

**Scattering spectra and corresponding electron microscope images of bowties with a single QD in their gaps.** Red solid lines are fits to the coupled oscillator model. For these fits we fixed the value of the plasmon linewidth to the average value measured from empty bowties, 0.385 eV. The coupled oscillator model does not fit the two bottom spectra well, but the structure of the spectra and their widths clearly point to splitting (compare with Supplementary Figure 1).

**Supplementary Figure 4**

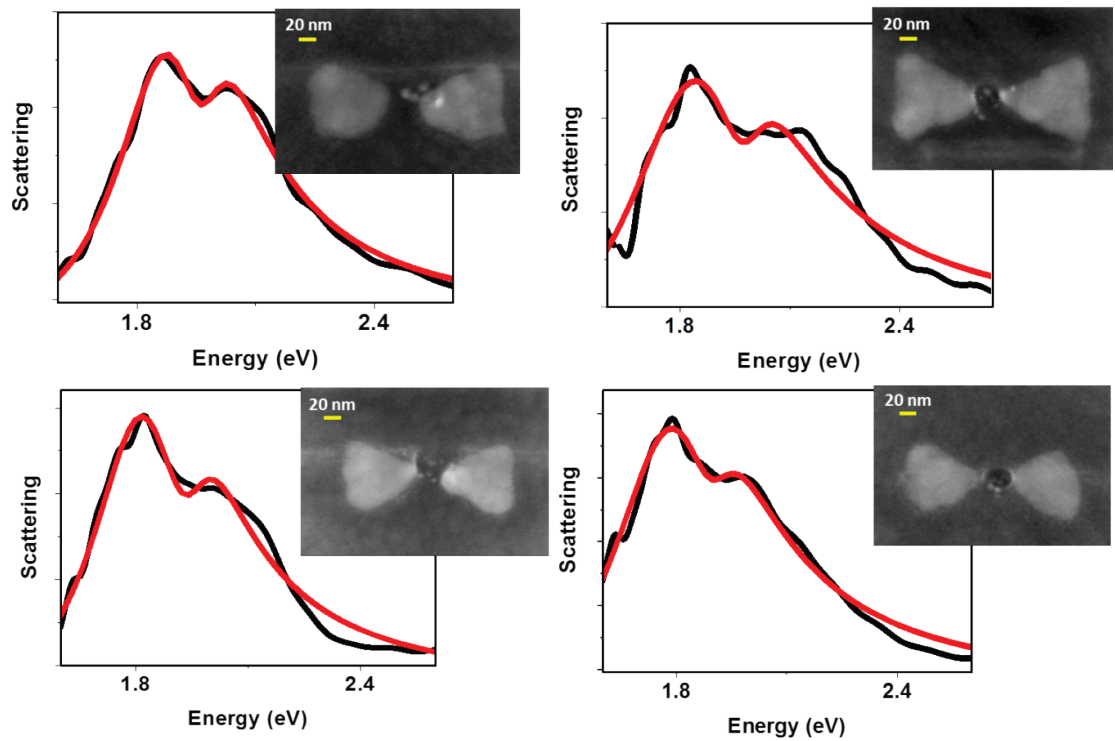

**Scattering spectra of bowties with several QDs within the gap.** The insets show scanning electron microscope images of the bowties. The red solid lines are fits to the coupled oscillator model.

**Supplementary Figure 5**

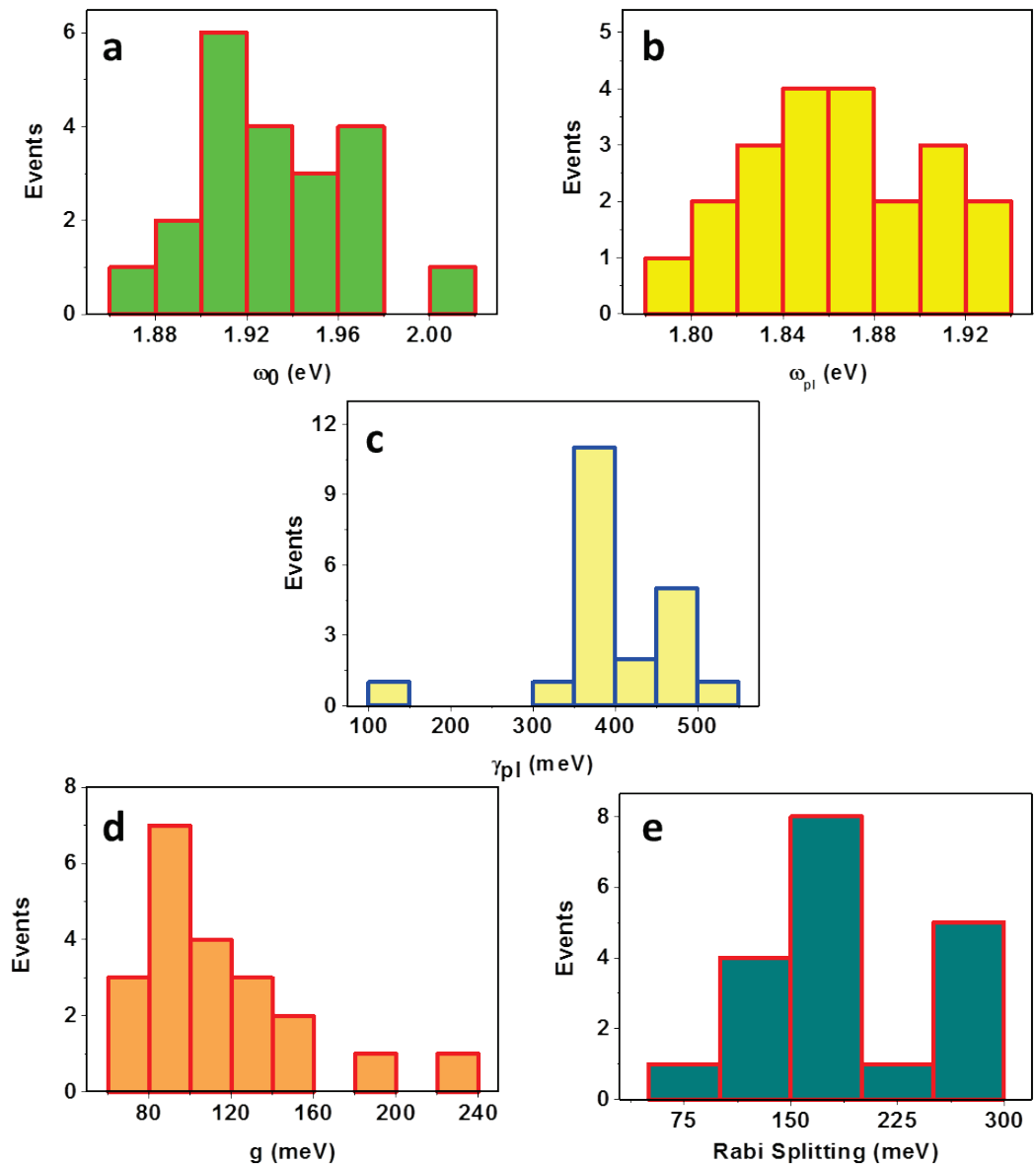

**Parameters from coupled oscillator model fits to scattering spectra.** The figure shows the fitting parameters obtained from fits of scattering spectra to the coupled oscillator model (equation 1 in Online Methods). **a.** Resonance frequency of the quantum emitter(s). **b.** Plasmon resonance frequencies. **c.** Plasmon linewidths. **d.** Coupling rates. **e.** Rabi splittings calculated directly from the fitted spectra.

**Supplementary Figure 6**

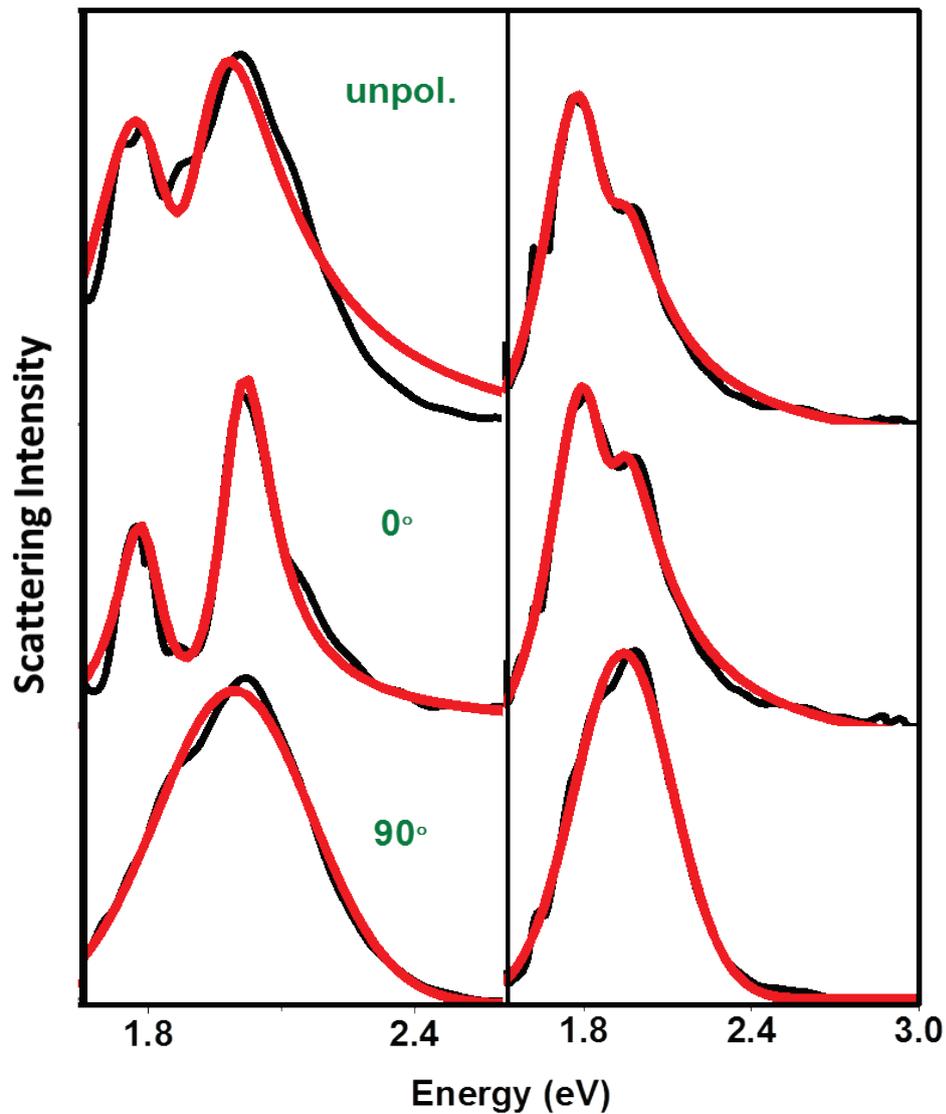

**Polarization dependence of scattering spectra.** Here two experiments similar to Fig. 2d of the main text are shown, with four QDs (left) and two QDs (right). Rabi splitting is observed even with unpolarized light (top row), but is strongest when the laser polarization is parallel to the bowtie long axis (0°, middle row). The splitting vanishes when the polarization is rotated by 90° (bottom row).

**Supplementary Figure 7**

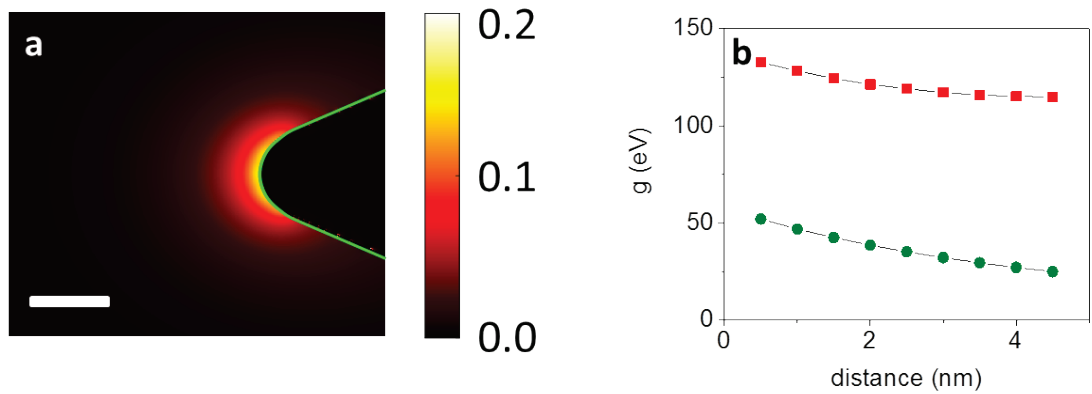

**Electromagnetic simulations of coupling. a.** The distribution of the coupling rate (in eV) of a quantum emitter with an oscillator strength of 0.6 near a single prism. The white bar represents 10 nm. **b.** Distribution of the coupling rate as a function of distance from one of the prisms of a bowtie (red) or from a single prism (green), along the center line and starting at the position where the edge of an 8 nm QD is 0.5 nm away from the prism.

**Supplementary Figure 8**

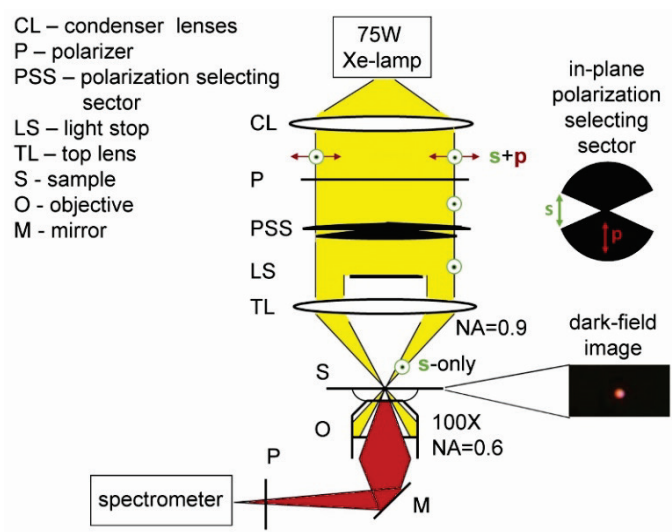

**Schematic of the dark-field microspectromter.** The setup was based on an inverted microscope equipped with a dark field condenser (NA=0.9), a 75 W Xenon lamp (Olympus) and a 100× oil immersion objective (NA=0.6). A combination of a light-stop with the top-lens of the condenser was used to exclude the light propagating along the normal to the sample plane, such that the light rays impinged on the sample at an angle of 70 degrees. A SpectraPro-150 spectrograph with a 1200 g/mm grating (Acton) and a Newton spectroscopy CCD camera (Andor Technology) were used to disperse the scattered light and register spectra. Polarization of the excitation light was controlled by a polarizer and an additional light-stop that passed the light only through a narrow selecting sector. When the sector was aligned such that the axis of the polarizer was normal to the bisecting line of the sector, the light at the sample was predominantly s-polarized.

**Supplementary Note**

The absorption spectrum of a semiconductor QD (Supplementary Figure 2) does not have a simple line shape like the emission spectrum, due to the absorption into higher energy states. How does that affect the coupling process? A good way to think about this problem is to envision this process in the time domain. First, a photon is injected into the cavity. As this photon bounces in the cavity it interacts with the quantum emitter. If the photon energy is in resonance with the emitter it absorbs the photon. Indeed, a QD contains more than a single exciton state. Each of the states that overlap the photon energy can absorb the photon. This photon has to be re-emitted. Typically a fast relaxation then brings the QD to the lowest-energy exciton, so the emission is from that exciton. However, irrespective of which exciton emits, unless the emission spectrum overlaps strongly the cavity spectrum the photon will escape the cavity. If there is such overlap the photon stays in the cavity, and the process can be repeated more than one time, which is the essence of strong coupling.

The bottom line is that a prerequisite for strong coupling is good overlap between both the absorption and emission of the QD and the cavity. Whether one exciton state is involved or more should not matter, just as in the case of molecules the number of vibronic states involved does not seem to matter (a discussion of this issue for molecules is given in the review of Torma and Barnes, reference 9 of the manuscript, section 4.2). This issue might be approached through a detailed quantum calculation.